\documentclass[RNAAS]{aastex63}

\shorttitle{Simulated Solar Spectra}
\shortauthors{Gilbertson et al.}
\usepackage{hyperref}
\usepackage{graphicx}
\usepackage{amsmath}
\usepackage{gensymb}

\definecolor{cjg}{rgb}{0.0, 0.5, 0.5}
\definecolor{ebf}{rgb}{0.4, 0.0, 0.6}

\begin{document}

\correspondingauthor{Christian Gilbertson}
\email{cjg66@psu.edu}

\title{Towards Extremely Precise Radial Velocities: I. Simulated Solar Spectra for Testing Exoplanet Detection Algorithms}

\author[0000-0002-1743-3684]{Christian Gilbertson}
\affiliation{Department of Astronomy \& Astrophysics, The Pennsylvania State University, 525 Davey Lab, University Park, PA 16802, USA}
\affiliation{Center for Exoplanets \& Habitable Worlds, 525 Davey Laboratory, The Pennsylvania State University, University Park, PA 16802, USA}
\affiliation{Institute for Computational \& Data Sciences, The Pennsylvania State University}
\affiliation{Penn State Astrobiology Research Centers}

\author[0000-0001-6545-639X]{Eric B. Ford}
\affiliation{Department of Astronomy \& Astrophysics, The Pennsylvania State University, 525 Davey Lab, University Park, PA 16802, USA}
\affiliation{Center for Exoplanets \& Habitable Worlds, 525 Davey Laboratory, The Pennsylvania State University, University Park, PA 16802, USA}
\affiliation{Institute for Computational \& Data Sciences, The Pennsylvania State University}
\affiliation{Penn State Astrobiology Research Centers}

\author[0000-0002-9332-2011]{Xavier Dumusque}
\affiliation{Astronomy Department of the University of Geneva, 51 Chemin des Maillettes, 1290 Versoix, Switzerland}

\date{\today}

\section{Introduction}  \label{sec:intro}
Recent and upcoming stabilized spectrographs are pushing the frontier for Doppler spectroscopy to detect and characterize low-mass planets.
Specifications for these instruments are so impressive that intrinsic stellar variability is expected to limit their Doppler precision for most target stars \citep{Fischer2016}.
To realize their full potential, astronomers must develop new strategies for distinguishing true Doppler shifts from intrinsic stellar variability.  
Stellar variability due to star spots, faculae and other rotationally-linked variability are particularly concerning, as the stellar rotation period is often included in the range of potential planet orbital periods.
To robustly detect and accurately characterize low-mass planets via Doppler planet surveys, the exoplanet community must develop statistical models capable of jointly modeling planetary perturbations and intrinsic stellar variability.  
Towards this effort, this note presents simulations of extremely high resolution, solar-like spectra created with SOAP 2.0 \citep{Dumusque2014} that includes multiple evolving star spots.
We anticipate this data set will contribute to future studies developing, testing, and comparing  statistical methods for measuring physical radial velocities amid contamination by stellar variability. 

\section{Simulations} \label{sec:simulations}

We have constructed an empirically-informed time series of simulated solar spectra using SOAP 2.0 \citep{Dumusque2014}. 
SOAP 2.0 was used to output time series of stellar spectra that would be observed for a Sun-like star with evolving spots. 
It constructs the simulated spectra by using the spots locations and sizes at each epoch to compute linear combinations of real, spatially resolved spectra of quiet and active regions of the Sun, accounting for both rotational Doppler shifts and limb darkening.
The spots differentially rotate 
with a period of 25.05 days at the equator and 35.0 days at the pole. 
We have simulated 50 timeseries of solar spectra with two observations per night for a year.  Each of the 730 spectra has a spectral resolution $R > 700000$. 
These optimistic assumptions allow future analyses to assess the effectiveness of new techniques in a best-case scenario.  One can add photon noise, degrade the spectral resolution, downsample the amount of observations, and/or inject planetary Doppler shifts to obtain a set of spectra representative of realistic planet surveys.

We generate star spot properties, largely following \citet{Borgniet2015}. 
The star spot temperature deficit in SOAP 2.0 was set to $\Delta T_{spot}=605
\text{K}$. 
%
%
The size distribution of spots was taken from a lognormal distribution described in \citet{Baumann2005} (\ref{eq:baumann}). 

\begin{equation}
  \dfrac{dN}{dA} = \dfrac{1}{(\langle A \rangle \sqrt{2 \pi  \ \sigma_a \  \text{log}(\sigma_a)}} \exp\left[-\dfrac{(\text{log}(A)-\text{log}(\langle A \rangle))^2}{2 \text{log}(\sigma_a)}\right]
  \label{eq:baumann}
\end{equation}
\noindent where $A$ is the maximum area of the spot in microsolar hemispheres (MSH)
$\text{log}(A)$ is normally distributed with $\mu = \text{log}(\langle A \rangle \ \sigma_a)$ and $\sigma = \sqrt{\text{log}(\sigma_a)}$. 
Following \citet{Borgniet2015}, 40\% of spots were labeled as ``isolated spots'' and the remaining were labeled as ``complex spot groups''. 
Isolated spots and complex spot groups have $\langle A \rangle = (46.51 \ \text{MSH}, 90.24 \ \text{MSH})$, and $\sigma_a = (2.14, 2.49)$, respectively  \citep{Borgniet2015}. 
Star spots smaller than 10 MSH and negelected as they impart RV signals that are insignificant compared to those from larger spots. 

Spot areas ($A(t)$) follow a parabolic decay law (\ref{eq:lifetime}).
\begin{equation}
  A(t) = \left(\sqrt{A_0}-\dfrac{D}{\sqrt{A_0}}(t-t_0)\right)^2
  \label{eq:lifetime}
\end{equation}
\noindent where $A_0$ is the maximum spot area, $D$ is the average decay rate, and $t_0$ is the time of maximum spot size.
%
%
The average decay rate distribution of spots was taken from a lognormal distribution described in \citet{Pillet1993} (\ref{eq:pillet}). 
\begin{equation}
  \dfrac{dN}{dD} = \dfrac{1}{D \sqrt{2 \pi} \ \sigma_p} \exp\left[-\dfrac{(\text{log}(D)-\mu_p)^2}{2 \sigma_p^2}\right]
  \label{eq:pillet}
\end{equation}
\noindent where $D$ is the average decay rate of the spot in MSH/day, $\mu_p$ is the log of the median of the distribution, and $\sigma_p$ is a measure of the width of the distribution. 
For isolated spots and complex spot groups respectively, $\mu_p = (2.619, 3.373)$ and $\sigma_b = (0.806, 0.869)$ \citep{Pillet1993}.  
Average decay rates were drawn between 3-200 MSH/day, as in \citet{Borgniet2015}. 
Spot lifetimes are calculated by dividing the maximum spot area by the average decay rate. 
Sunspots grow 10-11 times faster than they decay \citep{Howard1992} and this 
growth period is added in proportion to the decay lifetime. 

Spots were initialized at random times with a density set to 109 spots/stellar rotation period, consistent with the amount of spots greater than 10 MSH found by \citet{Meunier2010} in USAF/NOAA data (\href{http://www.ngdc.noaa.gov/stp/SOLAR/}{http://www.ngdc.noaa.gov/stp/SOLAR/}) from May 5, 1996 to October 7, 2007 ($\sim$ 1 solar cycle). 

The spot latitudes were drawn from a truncated normal distribution with mean 15.1$\degree$, width $\sigma=7.3\degree$ and limited to $\pm90\degree$  (\citet{Mandal2017}).
Spots were initialized randomly 
in longitude. 
The inclination of the stellar rotation axis is 90$\degree$ relative to the observer.

In addition to the solar-like data, we have constructed another 50 years of data for a similar level of activity of longer-lived spots by dividing the amount of spots created and spot decay rates by a factor of 30. 
Nine years of spectra for each type of activity are available online (doi:\href{https://doi.org/10.5281/zenodo.3753254}{10.5281/zenodo.162965} \citep{gilbertson_christian_2020_3753254})
The remaining 41 years are available by request from the authors.

\begin{figure}
\centering
\includegraphics[width=18cm]{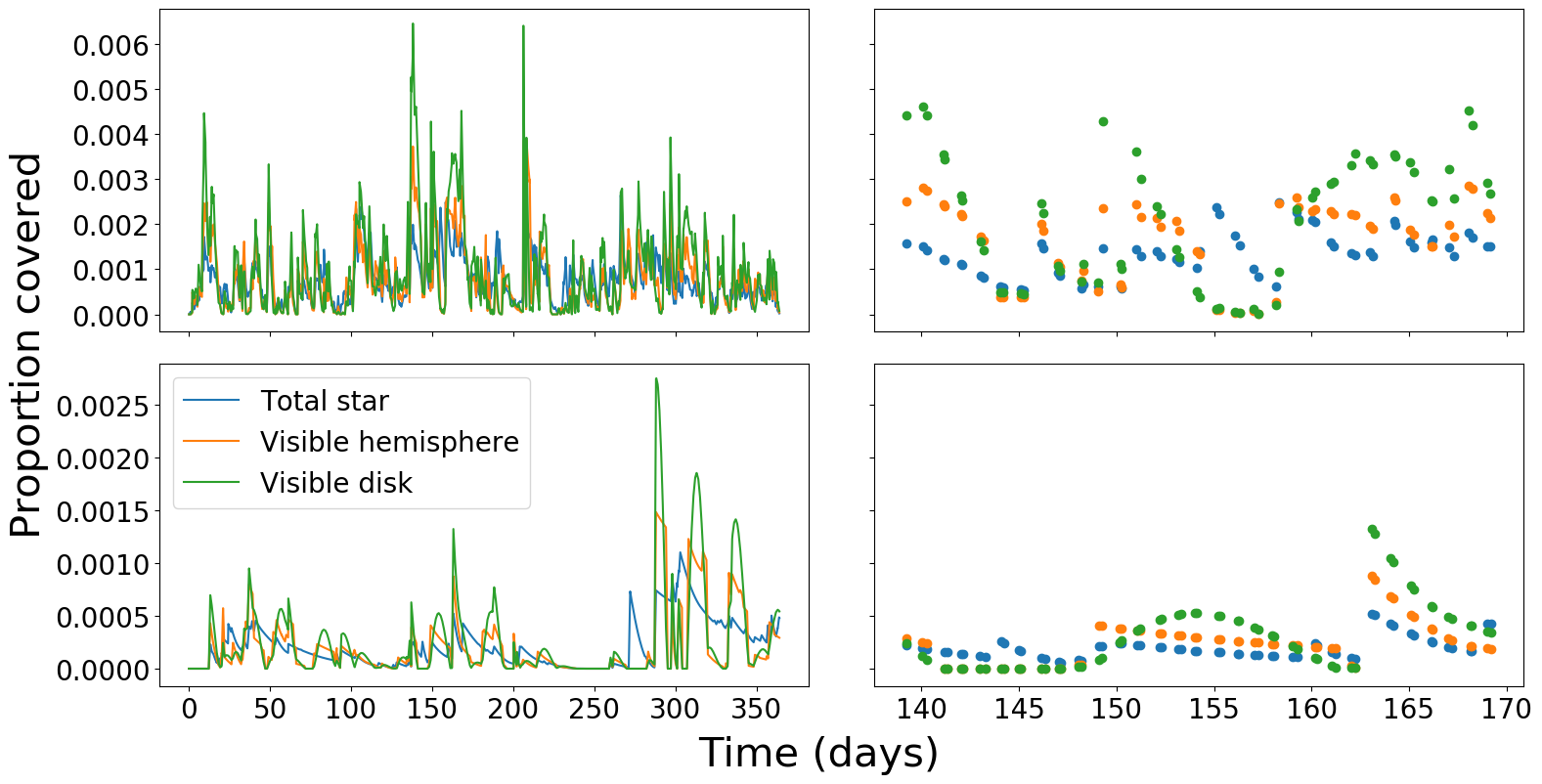}
\caption{The extent of star spots in representative simulations. The proportion of the total star (blue), visible hemisphere (orange), and visible disk (green) that are covered by spots at each epoch. The top row shows one realization for solar-like spot properties, while the bottom row shows a realization with longer-lived spots (see  \S\ref{sec:simulations}). The left column shows one year, and the right column zooms in on a single month.}
\label{fig:example}
\end{figure}

\section{Discussion} \label{sec:summary}
We present a set of empirically-informed solar spectra simulations that follow activity levels described in the literature, including realistic distributions for spot areas, lifetimes, latitudes, and rotation rates.  
This data set can be used to test and improve algorithms for mitigating the effects of stellar variability on Doppler planet searches.
Future work could incorporate 
additional viewing orientations, solar cycles, and/or additional resolved spectra used as inputs to SOAP 2.0.  

The authors wish to thank Ari Silburt for writing the initial versions of 
scripts used in this work. 
This work was funded in part by NSF AST award \#161086.  
We acknowledge the Institute for Computational and Data Sciences (\url{http://ics.psu.edu/}) at The Pennsylvania State University, including the CyberLAMP cluster supported by NSF grant MRI-1626251, for providing advanced computing resources and services that have contributed to the research results reported in this paper.
This project has received funding from the ERC under the European Union’s H2020 R\&I programme (grant No 851555).
\software{SOAP 2.0 \citep{Dumusque2014}}


\bibliography{bibliography}{}
\bibliographystyle{aasjournal}

\end{document}